\documentclass[10pt]{llncs}

\usepackage{epsfig,amsfonts,subfigure,amsbsy,bm,mathrsfs}
\usepackage{graphicx,cite,amssymb,amsmath,color}
\usepackage{amsmath,amssymb,amscd}
\usepackage{mathtools}
\usepackage[nolist]{acronym}
\usepackage{tabularx,booktabs}
\usepackage{multirow}
\usepackage{setspace}
\usepackage{xcolor}
\mathtoolsset{showonlyrefs=true}

\def\forcemath#1{\ifmmode #1 \else $#1$\fi}

\newenvironment{smallarray}[1]
 {\null\,\vcenter\bgroup\small
  \arraycolsep=.13885em
  \hbox\bgroup$\array{@{}#1@{}}}
 {\endarray$\egroup\egroup\,\null}

\newcolumntype{x}[1]{>{\centering\arraybackslash\hspace{0pt}}p{#1}}

\newcommand{\hh}{\noalign{\vskip 1.5mm} \hline \noalign{\vskip 1.5mm}}
\newcommand{\smallhline}{\noalign{\vskip .1mm} \hline \noalign{\vskip .4mm}}
\newcommand{\ccline}[1]{\noalign{\vskip 2mm} \cline{#1} \noalign{\vskip 2mm}}  
\newcommand{\mr}[2]{\multirow{ #1}{*}{#2}}

\newcommand{\B}{\forcemath{\bm{B}}}
\newcommand{\proto}{\mathscr{P}}

\newcommand{\BMA}[2]{\forcemath{{\left(#1 \,\, #2 \right)}}}

\newcommand{\BMsmallA}[2]{\forcemath{{\left(\!\begin{smallarray}{cc} #1 & #2 \end{smallarray}\!\right)}}}
\newcommand{\BMsmallB}[3]{\forcemath{{\left(\!\begin{smallarray}{ccc} #1 & #2 & #3 \end{smallarray}\!\right)}}}

\newcommand{\BMHWAsmallA}[4]{\forcemath{{\left(\begin{smallarray}{cc|cc}
	#1 & #2 & 1 & 0 \\
	#2 & #1 & 0 & 1 \\ \smallhline
	0  & 0  & #3 & #4 
	\end{smallarray}
	\right)}}}
\newcommand{\BMHWAsmallB}[6]{\forcemath{{\left(\begin{smallarray}{ccc|ccc}
	#1 & #2 & #3 & 1  & 0  & 0 \\
	#3 & #1 & #2 & 0  & 1  & 0 \\
	#2 & #3 & #1 & 0  & 0  & 1 \\ \smallhline
	0  & 0  & 0	 & #4 & #5 & #6 
	\end{smallarray}
	\right)}}}

\newcommand{\thrspa}{\forcemath{{\delta^\star_{\mathsf{SPA}}}}}
\newcommand{\thrtmp}{\forcemath{{\delta^\star_{\mathsf{TMP}}}}}

\newcommand{\ensemble}{\forcemath{{\mathscr{C}}}}
\newcommand{\ens}[1]{\forcemath{{\ensemble_{\mathsf{#1}}}}}

\newcommand{\vnt}{\forcemath{\mathsf{V}}}
\newcommand{\cnt}{\forcemath{\mathsf{C}}}
\newcommand{\degVN}{\ensuremath{d_v}}
\newcommand{\degCN}{\ensuremath{d_c}}
\newcommand{\degHWA}{\ensuremath{d_Q}}

\newcommand{\weight}[1]{\mathsf{wht}\left(#1 \right)}

\renewcommand{\vec}[1]{\ensuremath{\bm{#1}}}

\newcommand{\F}{\mathbb{F}}
\newcommand{\mat}[1]{\bm{#1}}
\newcommand{\sizeCirc}{p}
\newcommand{\Dec}[1]{\ensuremath{\text{DEC}_{#1}}} 

\begin{document}
	
\begin{acronym}
\acro{APP}{a posteriori probability}
\acro{BSC}{binary symmetric channel}
\acro{CN}{check node}
\acro{DE}{density evolution}
\acro{MDPC}{moderate-density parity-check}
\acro{QC}{quasi cyclic}
\acro{SPA}{sum-product algorithm}
\acro{BP}{belief propagation}
\acro{VN}{variable node}
\acro{CCA-2}{chosen ciphertext attack 2}
\acro{FER}{frame error rate}
\acro{LDPC}{low density parity-check}
\acro{PKC}{public-key cryptosystem}
\acro{HWA}{Hamming weight amplifier}
\acro{BF}{bit-flipping}
\acro{TMP}{ternary message passing}
\acro{BMP}{binary message passing}
\acro{RSA}{Rivest-Shamir-Adleman}
\acro{MP}{message passing}
\end{acronym}

\title{Protograph-Based Decoding of LDPC Codes with Hamming Weight Amplifiers}
\author{
Hannes Bartz\inst{1} \and 
Emna Ben Yacoub\inst{2} \and 
Lorenza Bertarelli\inst{3} \and 
Gianluigi Liva\inst{1}}
\institute{
Institute of Communication and Navigation, \\ German Aerospace Center (DLR), Wessling, Germany \\
\email{hannes.bartz@dlr.de, gianluigi.liva@dlr.de}
\and
Institute for Communications Engineering, \\ Technical University of Munich, Munich, Germany \\
\email{emna.ben-yacoub@tum.de}
\and
JMA wireless, Bologna, Italy \\
\email{lbertarelli@jmawireless.com}
}

\maketitle

\begin{abstract}
A new protograph-based framework for \ac{MP} decoding of \ac{LDPC} codes with \acp{HWA}, which are used e.g. in the NIST post-quantum crypto candidate LEDAcrypt, is proposed.  
The scheme exploits the correlations in the error patterns introduced by the \ac{HWA} using a turbo-like decoding approach where messages between the decoders for the outer code given by the \ac{HWA} and the inner \ac{LDPC} code are exchanged.
Decoding thresholds for the proposed scheme are computed using \ac{DE} analysis for \ac{BP} and \ac{TMP} decoding and compared to existing decoding approaches.
The proposed scheme improves upon the basic approach of decoding \ac{LDPC} code from the amplified error and has a similar performance as decoding the corresponding \ac{MDPC} code but with a significantly lower computational complexity.
\end{abstract}

\acresetall
\section{Introduction}\label{sec:intro}

In 1978, McEliece proposed a code-based~\ac{PKC}~\cite{mceliece1978public} that relies on the hardness of decoding an unknown linear error-correcting code.
Unlike the widely-used~\ac{RSA} cryptosystem~\cite{RSA1978}, the McEliece cryptosystem is resilient against attacks performed on a quantum computer\ and thus is considered as~\emph{post-quantum} secure.
One drawback of the McEliece cryptosystem compared to the \ac{RSA} cryptosystem is the large key size and the rate-loss .
Many variants of the McEliece cryptosystem based on different code families were considered in the past.
In particular, McEliece cryptosystems based on \ac{LDPC} allow for very small keys but suffer from feasible attacks on the low-weight dual code~\cite{monico2000using}.

A variant based on \ac{QC}-\ac{LDPC} codes that uses a sparse column scrambling matrix, a so-called~\ac{HWA}, to increase the density of the public code parity-check matrix was proposed in~\cite{BaldiHWA}.
However, unfortunate choices of the column scrambling matrix allow for structural attacks~\cite{otmani2010cryptanalysis}.
In~\cite{baldi2008new} a scheme that defeats the attack in~\cite{otmani2010cryptanalysis} by using dense row scrambling matrices and less structured column scrambling matrices was presented.
Optimized code constructions for the cryptosystem proposed in~\cite{baldi2008new} were presented in~\cite{baldi2013optimization}.
The ideas and results of~\cite{baldi2008new,baldi2013optimization} are the basis for the LEDcrypt~\cite{BaldiLEDAcrypt} \ac{PKC} and authentication schemes that are candidates at the current post-quantum cryptosystem standardization by NIST.

In LEDAcrypt~\cite{BaldiLEDAcrypt}, an variant of the bit-flipping\footnote{In~\cite{BaldiLEDAcrypt} and other literature the \ac{BF} decoder is referred to as ``Gallager’s \ac{BF}'' algorithm although it is different from the algorithm proposed by Gallager in~\cite{Gallager63:LDPC}.} decoder~\cite{Rudolph1967Majority}, called ``Q-decoder'', that exploits the correlation in the error patterns due to the \ac{HWA}, is used. 
The ``Q-decoder'' has the same error-correction performance as a \ac{BF} decoder for the corresponding \ac{MDPC} code but has a significantly lower computational complexity~\cite{baldi2013optimization}.

In this paper, the request for designing and analyzing improved decoders for \ac{LDPC} codes with \acp{HWA} from~\cite[Chapter~5]{BaldiLEDAcrypt} is considered. In particular, a new protograph-based decoding scheme for \ac{LDPC} codes with \acp{HWA} is presented.
The new scheme provides a turbo-like decoding framework, where information between the decoder of the outer rate-one code given by the \ac{HWA} and the decoder of the inner \ac{LDPC} codes, is exchanged.
The proposed framework allows to compare, analyze and optimize \ac{MP} decoding schemes for \ac{LDPC} codes with \acp{HWA}.

The \ac{DE} analysis for \ac{BP} and \ac{TMP} decoding shows, that the proposed protograph-based scheme has in general a similar error-correction capability as the corresponding \ac{MDPC} code under \ac{MP} decoding and improves upon the basic approach of decoding the amplified error using an \ac{LDPC} decoder.
For some parameters, the protograph-based scheme improves upon the corresponding \ac{MDPC} decoding approach while having a lower computational complexity due to the sparsity of the extended graph.
The gains in the error-correction capability predicted by \ac{DE} analysis are validated by Monte Carlo simulations under \ac{BP} and \ac{TMP} decoding.
\section{Preliminaries}

\subsection{Circulant Matrices}

Denote the binary field by $\F_2$ and let the set of $m\times n$ matrices over $\F_2$ be denoted by $\F^{m\times n}$.
The set of all vectors of length $n$ over $\F_2$ is denoted by $\F_2^n$.
Vectors and matrices are denoted by bold lower-case and upper-case letters such as $\vec{a}$ and $\mat{A}$, respectively.
A binary circulant matrix $\mat{A}$ of size $\sizeCirc$ is a $\sizeCirc\times \sizeCirc$ matrix with coefficients in $\F_2$ obtained by cyclically shifting its first row $\mat{a}=\left(a_0, a_1, \ldots, a_{\sizeCirc-1}\right)$ to right, yielding
\begin{equation}
\mat{A}=\left(\begin{array}{cccc}
a_0 & a_1 & \cdots & a_{\sizeCirc-1}\\
a_{\sizeCirc-1} & a_0 & \cdots & a_{\sizeCirc-2}\\
\vdots & \vdots & \ddots &\vdots\\
a_1 & a_2 & \cdots & a_{0}\\
\end{array}\right).
\end{equation} 
The set of $\sizeCirc\times \sizeCirc$ circulant matrices together with the matrix multiplication and addition forms a commutative ring and it is isomorphic to the polynomial ring $\left(\F_2[X]/\left(X^\sizeCirc-1\right),+,\cdot\right)$. 
In particular, there is a bijective mapping between a circulant matrix $\mat{A}$ and a polynomial $a(X)=a_0+a_1X+\ldots+a_{\sizeCirc-1}x^{\sizeCirc-1} \in \F_2[X]$. 
We indicate the vector of coefficients of a polynomial $a(X)$ as $\mat{a}=\left(a_0, a_1, \ldots, a_{\sizeCirc-1}\right)$. 
The weight of a polynomial $a(X)$ is the number of its non-zero coefficients, i.e., it is the Hamming weight of its coefficient vector $\mat{a}$. 
We indicate both weights with the operator $\weight{\cdot}$, i.e., $\weight{a(X)}=\weight{\mat{a}}$. 
In the remainder of this paper we use the polynomial representation of circulant matrices to provide an efficient description of the structure of the codes.

\subsection{McEliece Cryptosystem using LDPC Codes with Hamming Weight Amplifiers}

For $n=N_0\sizeCirc$, dimension $k=K_0\sizeCirc$, redundancy $r= n-k=R_0\sizeCirc$ with $R_0=N_0-K_0$ for some integer $\sizeCirc$, the parity-check matrix $\mat{H}(X)$ of a \ac{QC}-\ac{LDPC}\footnote{As in most of the literature, we loosely define a code to be \ac{QC} if there exists a permutation of its coordinates such that the resulting (equivalent) code has the following property: if $\mat{x}$ is a codeword, then any cyclic shift of $\mat{x}$ by $\ell$ positions is a codeword. For example, a code admitting a parity-check matrix as an array of $R_0\times N_0$ circulants does not fulfill the property above. However the code is \ac{QC} in the loose sense, since it is possible to permute its coordinates to obtain a code for which every cyclic shift of a codeword by $\ell=N_0$ positions yields another codeword.} code in polynomial form is a $R_0\times N_0$ matrix where each entry (polynomial) describes the corresponding circulant matrix.
We denote the corresponding $R_0\times N_0$ base matrix that indicates the Hamming weights of the polynomials in $\mat{H}(X)$ by
\begin{equation}
	\mat{B}_{H}=
	\begin{pmatrix}
	 b_{00} & b_{01} & \dots & b_{0(N_{0}-1)}
	 \\
	 b_{10} & b_{11} & \dots & b_{1(N_{0}-1)}
	 \\
	 \vdots & \vdots & \ddots & \vdots
	 \\
	 b_{(R_{0}-1)0} & b_{(R_{0}-1)1} & \dots & b_{(R_{0}-1)(N_{0}-1)}
	\end{pmatrix}.
\end{equation}

The column scrambling matrix $\mat{Q}(X)$ is of the form
\begin{equation}\label{eq:HWA}
	\mat{Q}(X)=
	\begin{pmatrix}
	 q_{00}(X) & \dots & q_{0(N_0-1)}(X)
	 \\
	 \vdots & \ddots & \vdots
	 \\
	 q_{(N_0-1)0}(X) & \dots & q_{(N_0-1)(N_0-1)}(X)
	\end{pmatrix}.
\end{equation}

We denote the corresponding base matrix for $\mat{Q}(X)$ by
\begin{equation}
	\mat{B}_{Q}
	=
	\begin{pmatrix}
	 b_{0} & b_{1} & \dots & b_{N_0-1}
	 \\
	 b_{N_0-1} & b_0 & \dots & b_{N_0-2}
	 \\
	 \vdots & \vdots& \ddots & \vdots
	 \\
	 b_{1} & b_{2}& \dots & b_0
	\end{pmatrix}
\end{equation}
where $\sum_{i=0}^{N_0-1}b_i=\degHWA$. 
This implies that $\mat{Q}(X)$ has constant row and column weight $\degHWA$.

Without loss of generality we consider in the following codes with $r=\sizeCirc$ (i.e. $R_0=1$). 
This family of codes covers a wide range of code rates and is of particular interest for cryptographic applications since the parity check matrices can be characterized in a very compact way.
The parity-check matrix of \ac{QC}-\ac{LDPC} codes with $r=\sizeCirc$ has the form 
\begin{equation}\label{eq:H_QC-LDPC-simple}
 \mat{H}(X)=
 \begin{pmatrix}
  h_{0}(X) & h_{1}(X) & \dots & h_{N_0-1}(X)
 \end{pmatrix}.
\end{equation}
Let $\Dec{\mat{H}}(\cdot)$ be an efficient decoder for the code defined by the parity-check matrix $\mat{H}$ that returns an estimate of a codeword or a decoding failure.

\medskip
\noindent
\textbf{Key generation:}
\begin{itemize}
 \item Randomly generate a parity-check matrix $\mat{H}\in\F_2^{r\times n}$ of the form \eqref{eq:H_QC-LDPC-simple} with $\weight{h_{i}(X)}=\degCN^{(i)}$ for $i=0,\dots,N_0-1$ and an invertible column scrambling matrix $\mat{Q}\in\F_2^{n\times n}$ of the form~\eqref{eq:HWA}.
 The matrix $\mat{H}$ with row weight $\degCN=\sum_{i=0}^{N_0-1}\degCN^{(i)}$ and the matrix $\mat{Q}$ with row and column weight $\degHWA$ is the \emph{private} key.

 \item From the private matrices $\mat{H}(X)$ and $\mat{Q}(X)$ the matrix $\mat{H}'(X)$ is obtained as
 \begin{equation}
 	\mat{H}'(X)=\mat{H}(X)\mat{Q}(X)=
 	\begin{pmatrix}
  		h'_{0}(X) & \dots & h'_{N_0-1}(X)
 	\end{pmatrix}.
 \end{equation}
 The row weight $\degCN'$ of $\mat{H}'(X)$ is upper bounded by $$\degCN'\leq\degCN\degHWA.$$
 Due to the low density of $\mat{H}$ and $\mat{Q}$ we have that $\degCN'\approx\degCN\degHWA$.
 Hence, the density of $\mat{H}'$ is higher compared to $\mat{H}$ which results in a degraded error-correction performance.
 Depending on $\degCN$ and $\degHWA$, $\mat{H}'$ may be a parity-check matrix of a \ac{QC}-\ac{MDPC} code~\cite{Misoczki13:MDPC}.
 
 The \emph{public} key is the corresponding binary $k\times n$ generator matrix $\mat{G}'(X)$ of $\mat{H}'(X)$ in systematic form.
\end{itemize}

\medskip
\noindent
\textbf{Encryption:}
\begin{itemize}
 \item To encrypt a plaintext\footnote{We assume that the CCA-2 security conversions from~\cite{kobara2001semantically} are applied to the McEliece cryptosystem to allow for systematic encoding without security reduction.} $\vec{u}\in\F_2^{k}$ a user computes the ciphertext $\vec{c}\in\F_2^n$ using the public key $\mat{G}'$ as
 \begin{equation}\label{eq:encryption}
  	\vec{c}=\vec{u}\mat{G}'+\vec{e}
  \end{equation} 
  where $\vec{e}$ is an error vector uniformly chosen from all vectors from $\F_2^{n}$ of Hamming weight $\weight{\mat{e}}= e$.
\end{itemize}

\medskip
\noindent
\textbf{Decryption:}
\begin{itemize}
 \item To decrypt a ciphertext $\vec{c}$ the authorized recipient uses the secret matrix $\mat{Q}$ to compute the transformed ciphertext
 \begin{equation}\label{eq:transformCiphertext}
 	\tilde{\vec{c}}=\vec{c}\mat{Q}^T=\vec{u}\mat{G}'\mat{Q}^{T}+\vec{e}\mat{Q}^{T}.
 \end{equation}
 A decoder $\Dec{\mat{H}}(\cdot)$ using the secret matrix $\mat{H}$ is applied to decrypt the transformed ciphertext $\tilde{\vec{c}}$ as
 \begin{equation}\label{eq:decryption}
  	\hat{\vec{c}}=\Dec{\mat{H}}(\tilde{\vec{c}})=\Dec{\mat{H}}(\vec{u}\mat{G}'\mat{Q}^{T}+\vec{e}\mat{Q}^{T}).
 \end{equation} 
 \item The generator matrix corresponding to $\mat{H}$ is used to recover the plaintext $\vec{u}$ from $\hat{\vec{c}}$.
\end{itemize}

\subsection{Protograph Ensembles}\label{subsec:ens}

A protograph $\proto$ \cite{Thorpe03:PROTO} is a small bipartite graph comprising a set of $N_0$ \acp{VN} (also referred to as \ac{VN} types) $\left\{\vnt_0, \vnt_1, \ldots, \vnt_{N_0-1}\right\}$ and a set of $M_0$ \acp{CN} (i.e., \ac{CN} types)  $\left\{\cnt_0, \cnt_1, \ldots, \cnt_{M_0-1}\right\}$. A \ac{VN} type $V_j$ is connected to a \ac{CN} type $\cnt_i$ by $b_{ij}$ edges. 
A protograph can be equivalently represented in matrix form by an $M_0\times N_0$ matrix $\B$. The $j$th column of $\B$ is associated to  \ac{VN} type $\vnt_j$ and the $i$th row of $\B$ is associated to \ac{CN} type $\cnt_i$. The $(i,j)$ element of $\B$ is  $b_{ij}$. 
A larger graph (derived graph) can be obtained from a protograph by applying a copy-and-permute procedure. The protograph is copied $Q$ times ($Q$ is commonly referred to as lifting factor), and the edges of the different  copies are permuted preserving the original protograph connectivity: If a type-$j$ \ac{VN} is connected to a type-$i$ \ac{CN} with $b_{ij}$ edges in the protograph, in the derived graph each type-$j$ \ac{VN} is connected to $b_{ij}$ distinct type-$i$ \acp{CN} (observe that multiple connections between a \ac{VN} and a \ac{CN} are not allowed in the derived graph). 
The derived graph is a Tanner graph with $n_0=N_0\sizeCirc$ \acp{VN} and $m_0=M_0\sizeCirc$ \acp{CN} that can be used to represent a binary linear block code.
A protograph $\proto$ defines a code ensemble $\ensemble$. 
For a given protograph $\proto$, consider all its possible derived graphs with $n_0=N_0\sizeCirc$ \acp{VN}. 
The ensemble $\ensemble$ is the collection of codes associated to the derived graphs in the set.

A distinctive feature of protographs is the possibility of specifying graphs which contain \acp{VN} which are associated to codeword symbols, as well as \acp{VN} which are not associated to codeword symbols. 
The latest class of \acp{VN} are often referred to as \emph{state} or \emph{punctured} \acp{VN}. 
The term ``punctured'' is used since the code associated with the derived graph can be seen as a punctured version of a longer code associated with the same graph for which all the \acp{VN} are associated to codeword bits. 
The introduction of state \acp{VN} in a code graph allows designing codes with a remarkable performance in terms  of error correction \cite{Divsalar04:ARA,Divsalar09:ProtoJSAC,liva_pexit}. 

\subsection{Decoding Algorithms for Low-Density Parity-Check Codes}

In this work we consider two types of \ac{MP} decoding algorithms for \ac{LDPC} codes.

\subsubsection{Scaled Sum-Product Algorithm}

We consider a \ac{BP} decoding algorithm that is a generalization of the classical \ac{SPA}, where the generalization works by introducing an attenuation of the extrinsic information produced at the \acp{CN} (see~\cite{bartz2019decoding} for details). 
As we shall see, the attenuation can be used as an heuristic method to control the code performance at low error rates where trapping sets may lead to premature error floors.  
% The channel message is initialized as 
% \[
% \mch=c \ln \frac{n-e}{n}
% \]
% for the variable nodes associated to a ciphertext bit (recall that $c=\pm1$ is the ciphertext bit), whereas $\mch=0$ for the state variable nodes.
% At the variable nodes,
% \begin{equation}
% \msg{\vn}{\cn}=\mch+\sum_{\cn'\in\neigh{\vn}\backslash\cn}{\msg{\cn'}{\vn}} \label{eq:SPA_VN}
% \end{equation}
% while at the check nodes
% \begin{equation}
% \msg{\cn}{\vn}=\omega 2\tanh^{-1}\left[ \prod_{\vn'\in\neigh{\cn}\backslash\vn}\tanh\left(\frac{\msg{\vn'}{\cn}}{2}\right)\right] \label{eq:SPA_CN}
% \end{equation}
% with final decision, after iterating \eqref{eq:SPA_VN}, \eqref{eq:SPA_CN} a given number of times, given by 
% \begin{equation}
% \hat{x}=\sign\left[\mch+\sum_{\cn\in\neigh{\vn}}{\msg{\cn}{\vn}}\right] \label{eq:SPA_final}.
% \end{equation}

\subsubsection{Ternary Message Passing (TMP)}

\ac{TMP} is an extension of \ac{BMP} decoding introduced in \cite{lechner_analysis_2012}. The exchanged messages between \acp{CN} and \acp{VN} belong to the ternary alphabet $\mathcal{M}=\{-1,0,1\}$, where $0$ corresponds to an erasure. At the \acp{CN} the outgoing message is the product of the incoming messages. The update rule at the \acp{VN} involves weighting the channel and the incoming \ac{CN} messages. 
The corresponding weights can be estimated from the \ac{DE} analysis (see~\cite{yacoub2019protograph}). 
A quantization function is then applied to map the sum of the weighted messages to the ternary message alphabet $\mathcal{M}$.
% Ternary message passing is a coarsly-quantized variant of \ac{SPA} decoding~\cite{yacoub2019protograph}.
% The channel message is initialized as 
% \[
% \mch=c \ln \frac{n-e}{n}
% \]
% for the variable nodes associated to a ciphertext bit (recall that $c=\pm1$ is the ciphertext bit), whereas $\mch=0$ for the state variable nodes.
% At the variable nodes,
% \begin{equation}
% \msg{\vn}{\cn}=f\left(\mch+\sum_{\cn'\in\neigh{\vn}\backslash\cn}{D_{C'V}\msg{\cn'}{\vn}}\right) \label{eq:TMP_VN}
% \end{equation}
% where 
% \[
% f(x)=\begin{cases}+1 & x>a \\ 0 & -a\leq x\leq a\\ -1 & x<-a\end{cases}
% \]
% for some quantization threshold $a$ and $D_{C'V}$.
% The value $a$ and the weights $D_{C'V}$ (for each iteration) can be pre-computed using the \ac{DE} analysis from~\cite{yacoub2019protograph}.
% At the check nodes
% \begin{equation}
% \msg{\cn}{\vn}= \prod_{\vn'\in\neigh{\cn}\backslash\vn}\msg{\vn'}{\cn} \label{eq:TMP_CN}
% \end{equation}
% with final decision, after iterating \eqref{eq:SPA_VN}, \eqref{eq:SPA_CN} a given number of times, given by 
% \begin{equation}
% \hat{x}=\sign\left[\mch+\sum_{\cn\in\neigh{\vn}}{D_{C'V}\msg{\cn}{\vn}}\right] \label{eq:TMP_final}.
% \end{equation}

\subsection{Decoding of \ac{QC}-\ac{LDPC} codes with Hamming Weight Amplifiers}

The decoding step in~\eqref{eq:decryption} using the parity-check matrix $\mat{H}$ is possible since $\vec{u}\mat{G}'\mat{Q}^T=\vec{x}'\mat{Q}^T$ is a codeword $\vec{x}$ of the \ac{LDPC} code $\mathcal{C}$ described by $\mat{H}$ since we have that
\begin{align}
 \vec{x}'\mat{H}'^T=\vec{x}'\mat{Q}^T\mat{H}^{T}=\mat{0}
 \quad\Longleftrightarrow\quad
 \vec{x}'\mat{Q}^T\in\mathcal{C}.
\end{align}

The error weight of transformed error $\vec{e}'=\vec{e}\mat{Q}^{T}$ in~\eqref{eq:decryption} is increased and upper bounded by
\begin{equation}
	\weight{\vec{e}'}\leq e\degHWA.
\end{equation}
Due to the sparsity of $\mat{H}$, $\mat{Q}$ and $\vec{e}$ we have that $\weight{\vec{e}'}\approx e\degHWA$.
In other words, the matrix $\mat{Q}$ increases the error weight and thus we call $\mat{Q}$ a \emph{\ac{HWA}}.
% Due to the transformation with $\mat{Q}$ the error locations (positions of the 1's) in $\vec{e}'$ are correlated.
% In particular, we can write the transformed error vector $\vec{e}'$ in terms of $\mat{Q}^T$ as
% \begin{equation}
% 	\vec{e}'=\sum_{i\in\supp{\vec{e}}}\vec{q}_i
% \end{equation}
% where $\vec{q_i}$ denotes the $i$-th row of $\mat{Q}^T$.
% Since the decoder knows $\mat{Q}$ the information about the correlated error positions can be incorporated to improve the error-correction performance of the decoder (see~\cite{BaldiLEDApkc,BaldiLEDAkemArXiv}).

In the following we consider two decoding principles for \ac{LDPC} codes with \acp{HWA}.

\subsubsection{Basic Decoding Approach}

A simple approach of decoding an~\ac{LDPC} code with~\ac{HWA} is to decode the transformed ciphertext $\vec{u}\mat{G}=\Dec{\mat{H}}(\tilde{\vec{c}})$ using a decoder for the~\ac{LDPC} code defined by $\mat{H}$ (see~\cite{baldi2008new,baldi2013optimization}).
Due to the sparse parity-check matrix $\Dec{\mat{H}}(\cdot)$ has a good error-correction performance but has to correct the amplified error $\vec{e}'$ of weight $\weight{\vec{e}'}\approx e\degHWA$.

\subsubsection{Decoding QC-LDPC-HWA Codes as QC-MDPC Codes}

An alternative decoding approach is to consider $\mat{H}'$ as a parity-check matrix of a \ac{QC}-\ac{MDPC} code and decode the ciphertext $\vec{c}$ without using the transformation in~\eqref{eq:transformCiphertext}:
\begin{equation}\label{eq:MDPCdec}
	\vec{u}\mat{G}'=\Dec{\mat{H}'}(\vec{c}).
\end{equation}
Compared to $\Dec{\mat{H}}(\cdot)$, the error-correction performance of $\Dec{\mat{H}'}(\cdot)$ is degraded due to the higher density of $\mat{H}'$.  
However, the decoder only has to correct errors of weight $\weight{\vec{e}}=e$ (instead of $e\degHWA$) only since the Hamming weight is \emph{not} increased by the transformation of the ciphertext $\vec{c}$ in~\eqref{eq:transformCiphertext}.

\subsubsection{Comparison of Decoding Strategies}

In order to evaluate the performance of the previously described decoding strategies for~\ac{LDPC} codes with~\ac{HWA}, we analyze the error-correction capability using~\ac{DE} analysis.
The analysis, which addresses the performance of the relevant code ensembles in the asymptotic regime, i.e., as $n$ goes to infinity, can be used to estimate the gains achievable in terms of error correction capability.
We denote the iterative decoding threshold under \ac{SPA} by $\thrspa$ and the decoding threshold of \ac{TMP} by $\thrtmp$. 
For a fair comparison, we consider the \ac{QC}-\ac{MDPC} ensemble for 80-bit security from~\cite{Misoczki13:MDPC} as a reference.
For the reference ensemble the estimate of the error-correction capability for a given $n$ under \ac{SPA} and \ac{TMP} decoding can be roughly obtained as $n\thrspa$ and $\thrtmp$, respectively.

The parameters of the corresponding \ac{LDPC} codes and the \acp{HWA} are chosen such that the row-weight of the resulting parity-check matrices $\mat{H}'=\mat{H}\mat{Q}$ match with the reference ensemble, i.e., we have $\degCN'=\degCN\degHWA=90$.
In this setting, the rough estimate of the error-correction capability of the basic decoder is obtained by $n\thrspa/\degHWA$ and $n\thrtmp/\degHWA$, where $\thrspa$ and $\thrtmp$ are the decoding thresholds of the corresponding \ac{LDPC} code under \ac{SPA} and \ac{TMP} decoding, respectively.

The results in Table~\ref{tab:thresholdsLDPCvsMDPC} show, that decoding the \ac{MDPC} code ($\degHWA=1$) gives a better error-correction performance than decoding the \ac{LDPC} code from the amplified error.

\begin{table}[ht!]
	\begin{center}
		\caption{Comparison of decoding thresholds for basic (\ac{LDPC}) and \ac{MDPC} decoding.}\label{tab:thresholdsLDPCvsMDPC}
		\begin{tabular}{cx{1.5cm}x{1.5cm}x{2.5cm}x{2.5cm}}
			\hline\hline
			Base matrix     & $n$   & $\degHWA$  & $n\thrspa/\degHWA$ & $n\thrtmp/\degHWA$  \\ \hline\hline\\[-7pt]
			\BMA{45}{45}    & $9602$ & $1$  & $113$ & $113$ \\	\hh	
			\BMA{15}{15}    & $9602$ & $3$	& $99$ & $89$\\	\hh
			\BMA{9}{9}      & $9602$ & $5$	& $87$ & $78$\\	\hh
			\BMA{5}{5}      & $9602$ & $9$  & $72$ & $62$\\	\hh
		\end{tabular}
	\end{center}
\end{table}

There is a bit-flipping-based~\cite{Rudolph1967Majority} decoder, called ``Q-decoder'', that incorporates the knowledge of the \ac{HWA} matrix $\mat{Q}$ during the decoding process~\cite{BaldiLEDAcrypt}.
The Q-decoder is equivalent to the bit-flipping decoder of the corresponding \ac{MDPC} code~\cite[Lemma~1.5.1]{BaldiLEDAcrypt} but has a significantly lower computational complexity~\cite[Lemma~1.5.2]{BaldiLEDAcrypt}.

\section{Improved Protograph-Based Decoding of LDPC Codes with Hamming Weight Amplifiers}

Motivated by the observations above, we derive a new protograph-based decoding framework for \ac{LDPC} codes with \acp{HWA} that incorporates the knowledge about the \ac{HWA} matrix $\mat{Q}$ at the receiver. 
The decoding framework allows to apply known \ac{MP} decoding algorithms and has an improved error-correction capability compared to the naive approach and a significantly reduced computational complexity compared to the corresponding \ac{MDPC} decoding approach (see~\eqref{eq:MDPCdec}).

\subsection{Protograph Representation of the LDPC-HWA Decoding Problem}
Let $\mathcal{C}$ be an \ac{LDPC} code with parity-check matrix $\mat{H}$ and let $\mathcal{C}'$ denote the code with parity-check matrix
\begin{equation}\label{eq:H_MDPC}
  \mat{H}'=\mat{H}\mat{Q}.
\end{equation}

Then we have that
\begin{equation}
 \vec{x}\mat{H}^{T}=\vec{0},\quad \forall \vec{x}\in\mathcal{C}
 \qquad\text{and}\qquad
 \vec{x}'\mat{H}'^{T}=\vec{0},\quad \forall \vec{x}'\in\mathcal{C}'.\label{eq:pce_MDPC}
\end{equation}

Using~\eqref{eq:H_MDPC} we can rewrite~\eqref{eq:pce_MDPC} as
\begin{align}\label{eq:parCheckMDPC}
\vec{x}'\mat{H}'^{T}=\vec{x}'(\mat{H}\mat{Q})^{T}={\vec{x}'\mat{Q}^{T}}\mat{H}^{T}=\vec{0}, \quad\forall \vec{x}'\in\mathcal{C}'.
\end{align}
Hence, we have that $\vec{x}'\mat{Q}^T$ must be contained in $\mathcal{C}$ for all $\vec{x}'\in\mathcal{C}'$.
Defining $\vec{x}=\vec{x}'\mat{Q}^T$ we can restate~\eqref{eq:pce_MDPC} as
\begin{align}
	\vec{x}&=\vec{x}'\mat{Q}^T 
	\\
	\vec{x}\mat{H}&=\vec{0} 
\end{align}
which we can write as $\left(\vec{x}' \ \vec{x}\right)\mat{H}_\text{ext}^{T}=\vec{0}$ with
\begin{equation}\label{eq:H_ext}
	\mat{H}_\text{ext}=
	\begin{pmatrix}
	 \mat{Q} & \mat{I}_{n\times n}
	 \\
	 \mat{0} & \mat{H}
	\end{pmatrix}.
\end{equation}
The matrix $\mat{H}_\text{ext}$ in~\eqref{eq:H_ext} is a $(n+r)\times 2n$ parity-check matrix of an \ac{LDPC} code of length $2n$ and dimension $n-r$. 
The corresponding base matrix of $\mat{H}_\text{ext}$ is
\begin{equation}\label{eq:B_ext}
	\mat{B}_\text{ext}=
	\begin{pmatrix}
	 \mat{B}_Q & \mat{I}
	 \\
	 \mat{0} & \mat{B}_H
	\end{pmatrix}.
\end{equation}

The extended parity-check $\mat{H}_\text{ext}$ can be used for decoding where the $n$ rightmost bits are associated to the punctured \acp{VN} and the $n$ leftmost bits are associated to the ciphertext $\vec{c}'$.
As mentioned in Section~\ref{subsec:ens} the introduction of state \acp{VN} in a code graph can improve the error-correction capability of the code significantly \cite{Divsalar04:ARA,Divsalar09:ProtoJSAC,liva_pexit}. 

\begin{figure}
    \centering
    \includegraphics[width=.8\columnwidth]{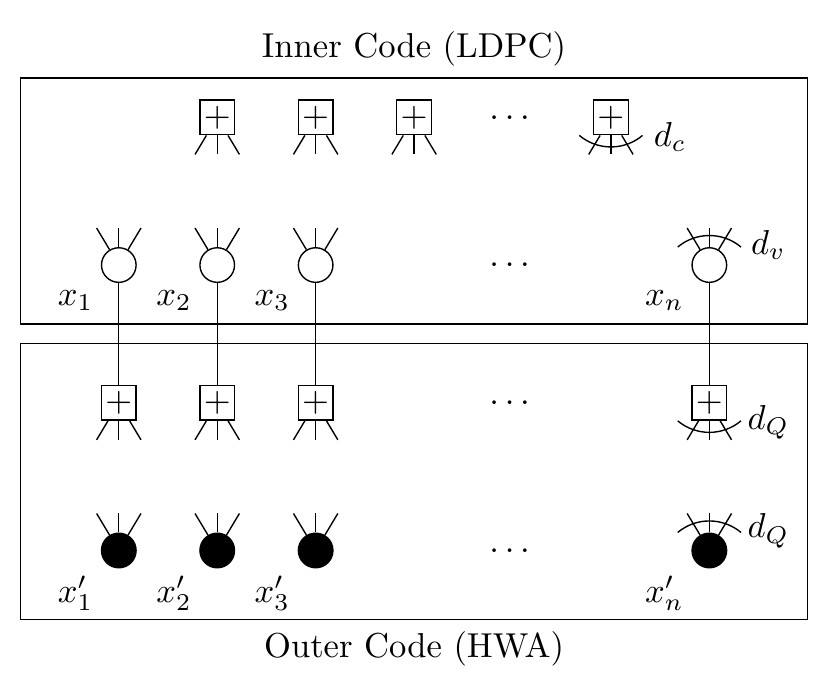}
    \vspace*{-10pt}
    \caption{Protograph representation of the \ac{LDPC} code with \ac{HWA}.}
    \label{fig:proto_HWA}
    \vspace*{-10pt}
\end{figure}

The protograph corresponding to the base matrix $\mat{B}_\text{ext}$ is depicted in Figure~\ref{fig:proto_HWA}. 

\subsection{Complexity Considerations}

The complexity of \ac{MP} decoding depends on the \ac{CN} and \ac{VN} degrees of the underlying graph.
Hence, the complexity of decoding the LDPC code is significantly lower than the complexity of decoding the corresponding \ac{MDPC} code (see e.g.~\cite{baldi2013optimization}).
For the protograph-based decoding approach, we have $\mathcal{O}\left((\degCN+\degHWA) n\right)$ $\ac{CN}$ and $\ac{VN}$ messages per iteration whereas in the corresponding \ac{MDPC} approach we have $\mathcal{O}\left(\degVN\degHWA n\right)$ $\ac{CN}$ and $\ac{VN}$ messages per iteration.
Hence, the protograph-based decoding approach has a significantly lower complexity compared to the \ac{MDPC} decoding approach.
\begin{example}
 This effect was also observed in a Monte Carlo simulation for the first ensemble in Table~\ref{tab:LEDAcryptParam}. 
 The simulation of $2\cdot10^4$ iterations with a non-optimized ANSI C implementation of the \ac{TMP} decoder took 12m 18s ($36.9\cdot10^{-3}$ $s/$iteration) for the protograph-based approach and 5h 50m 4s ($1.1$ $s/$iteration) for the corresponding \ac{MDPC} approach.
\end{example}

\section{Density Evolution Analysis}

We now provide an asymptotic analysis of the code ensembles resulting from the different decoding approaches for \ac{LDPC} codes with \acp{HWA}. 
The analysis is performed by means of \ac{DE} under \ac{BP} (\ac{SPA}) and \ac{TMP} decoding in order to get a rough estimate of the error correction capability of the codes drawn from the proposed ensembles.

\subsection{BP: Quantized Density Evolution for Protographs}

For \ac{BP}, we resort to quantized \ac{DE} (see~\cite{Chung01:DE,Jin2006:QDE} for details).
The extension to protograph ensembles is straightforward and follows the footsteps of~\cite{liva_pexit,Liva2013:Proto_TWCOM}. 
Simplified approaches based on the Gaussian approximation are discarded due to the large \ac{CN} degrees~\cite{chung2001analysis} of the \ac{MDPC} ensembles.

\subsection{TMP: Density Evolution for Protographs}

The decoding threshold $\thrtmp$, the optimal quantization threshold and the weights for the \ac{CN} messages for \ac{TMP} can be obtained by the \ac{DE} analysis in~\cite{yacoub2019protograph}. 

\subsection{Estimation of the Error Correction Capability}

In order to evaluate the error correction performance of the above described decoding scheme we compare the proposed protograph ensembles described by~\eqref{eq:B_ext} with the corresponding \ac{QC}-\ac{MDPC} ensemble.
As a reference we take the \ac{MDPC} ensemble $\mat{B}_\text{MDPC}=\left(45 \ 45\right)$ for $80$-bit security from~\cite{Misoczki13:MDPC}.

For a fair comparison the reference ensembles in Table~\ref{tab:thresholds} are designed such that the base matrix $\mat{B}$ of $\mat{H}\mat{Q}$ equals the base matrix $\mat{B}_\text{MDPC}$ of the corresponding \ac{QC}-\ac{MDPC} ensemble, i.e., we have
\begin{equation}
	\mat{B}_{H}\mat{B}_{Q}\overset{!}{=}\mat{B}_\text{MDPC}.
\end{equation}

For each ensemble, we computed the iterative decoding threshold, i.e., the largest channel error probability for which, in the limit of large $n$, \ac{DE} predicts successful decoding convergence.
We denote the iterative decoding threshold under \ac{SPA} by $\thrspa$ and the decoding threshold of \ac{TMP} by $\thrtmp$. 
In Table \ref{tab:thresholds} we provide a rough estimate of the number of errors at which the waterfall region of the block error probability is expected to be, by multiplying the asymptotic thresholds with the block length $n$. 

\begin{table}[ht!]
	\begin{center}
		\caption{Thresholds computed for different protographs.}\label{tab:thresholds}
		\begin{tabular}{ccx{1.5cm}x{1.5cm}x{1.5cm}}
			\hline\hline
			Ensemble & Base matrix & $n$ & $n\thrspa$ & $n\thrtmp$  \\ \hline\hline\\[-7pt]
			$\ens{A}$ & \BMsmallA{45}{45} 		& $9602$  	& $113$ & $113$ \\	\hh	
			$\ens{B}$ & \BMHWAsmallA{2}{1}{15}{15} & $9602$  	& $121$ & $103$\\	\hh
			$\ens{C}$ & \BMHWAsmallA{3}{2}{9}{9} & $9602$  	& $126$ & $101$\\	\hh
			$\ens{D}$ & \BMHWAsmallA{5}{4}{5}{5} & $9602$  	& $127$ & $80$\\	\hh
		\end{tabular}
	\end{center}
\end{table}

\subsection{Estimation of the Error Correction Capability for the LEDAcrypt Code Ensembles}

Table~\ref{tab:LEDAcryptParam} shows the decoding thresholds under \ac{SPA} and \ac{TMP} decoding for the protograph ensembles corresponding to the parameters in the current LEDAcrypt specifications~\cite[Table~3.1]{BaldiLEDAcrypt}. 
The decoding thresholds for the \ac{MDPC} ensembles under \ac{SPA} decoding could not be obtained by the quantized \ac{DE} due to the high \ac{CN} degrees. 
The results show, that the proposed protograph-based approach has a similar error correction capability as the corresponding \ac{MDPC} code.
Further, the error correction capability under \ac{SPA} and the efficient \ac{TMP} decoding significantly improves upon the basic decoding approach~\cite{baldi2008new,baldi2013optimization}.

\begin{table}[ht!]
	\begin{center}
	\caption{Thresholds for different protographs for the parameters of LEDAcrypt~\cite[Table~3.1]{BaldiLEDAcrypt} for the NIST categories 1 (128 Bit), 3 (192 Bit) and 5 (256 Bit).}\label{tab:LEDAcryptParam}
		\begin{tabular}{ccccccccc}
		\hline
		\hline
		\mr{2}{SL[Bit]} & \mr{2}{$\mat{B}_{ext}$} & \mr{2}{$\mat{B}_{MDPC}$} & \multirow{2}{*}{$n$} & \multicolumn{2}{c}{Proto} & \multicolumn{1}{c}{MDPC} & Basic
		\\ \cline{5-8}
		& & & & $n\thrspa$ & $n\thrtmp$ & $n\thrtmp$ & n\thrspa/\degHWA \\
		\hline
		\hline\noalign{\vskip 2mm}
		% Category 1 (128 Bit)
		\multirow{2}{*}[-0.5cm]{128} &  $\BMHWAsmallA{4}{3}{11}{11}$ & $\BMsmallA{77}{77}$ & $29878$ & 239 &  203 & 227 & 167
		\\ \ccline{2-8}
		& $\BMHWAsmallB{4}{3}{2}{9}{9}{9}$ & $\BMsmallB{81}{81}{81}$ & 23559 & 116  & 98 &  110 & 72
		% \\ \ccline{2-10}
		% & $\BMHWAsmallC{2}{2}{2}{1}{13}{13}{13}{13}$ & $\BMsmallC{91}{91}{91}{91}$ & $30188$ & 69 & 92  & 84  & --- & 93 & 66 
		\\ \hh

		% Category 3 (192 Bit)
		\multirow{2}{*}[-0.5cm]{192} & \BMHWAsmallA{5}{3}{13}{13} & \BMsmallA{104}{104} & 51386 & 316 & 272 & 303 & 222
		\\ \ccline{2-8}
		& $\BMHWAsmallB{4}{4}{3}{11}{11}{11}$ & $\BMsmallB{121}{121}{121}$ & $48201$ & 162 & 144  & 163 & 106
		% \\ \ccline{2-10}
		% & $\BMHWAsmallC{3}{2}{2}{2}{15}{15}{15}{15}$ & $\BMsmallC{135}{135}{135}{135}$ & 57364 & 101	& 120  & 114 & ---  & 126 & 87 
		\\\hh

		% Category 5 (256 Bit)
		\multirow{2}{*}[-0.5cm]{256} & \BMHWAsmallA{7}{6}{11}{11} & \BMsmallA{143}{143} & $73754$ & 344  &  302 & 331 & 223
		\\ \ccline{2-8}
		& $\BMHWAsmallB{4}{4}{3}{15}{15}{15}$ & $\BMsmallB{165}{165}{165}$  & 82311 & 205 & 189  & 214 & 144
		% \\ \ccline{2-10}
		% & $\BMHWAsmallC{4}{3}{3}{3}{13}{13}{13}{13}$ & $\BMsmallC{169}{169}{169}{169}$ & 90764 & 134 & 159 & 154 & --- & 172 & 107 
		\\\hh
		\end{tabular}
	\end{center}
\end{table}
\section{Simulation Results}

In order to evaluate the actual error correction capability and validate the gains predicted by \ac{DE} analysis, we performed Monte Carlo simulations for codes picked from the ensembles in Table~\ref{tab:thresholds}. 
The number of iterations for the \ac{SPA} and \ac{TMP} algorithm has been fixed to $100$.

\begin{figure}[ht!]
	\centering
	\includegraphics[width=0.68\columnwidth]{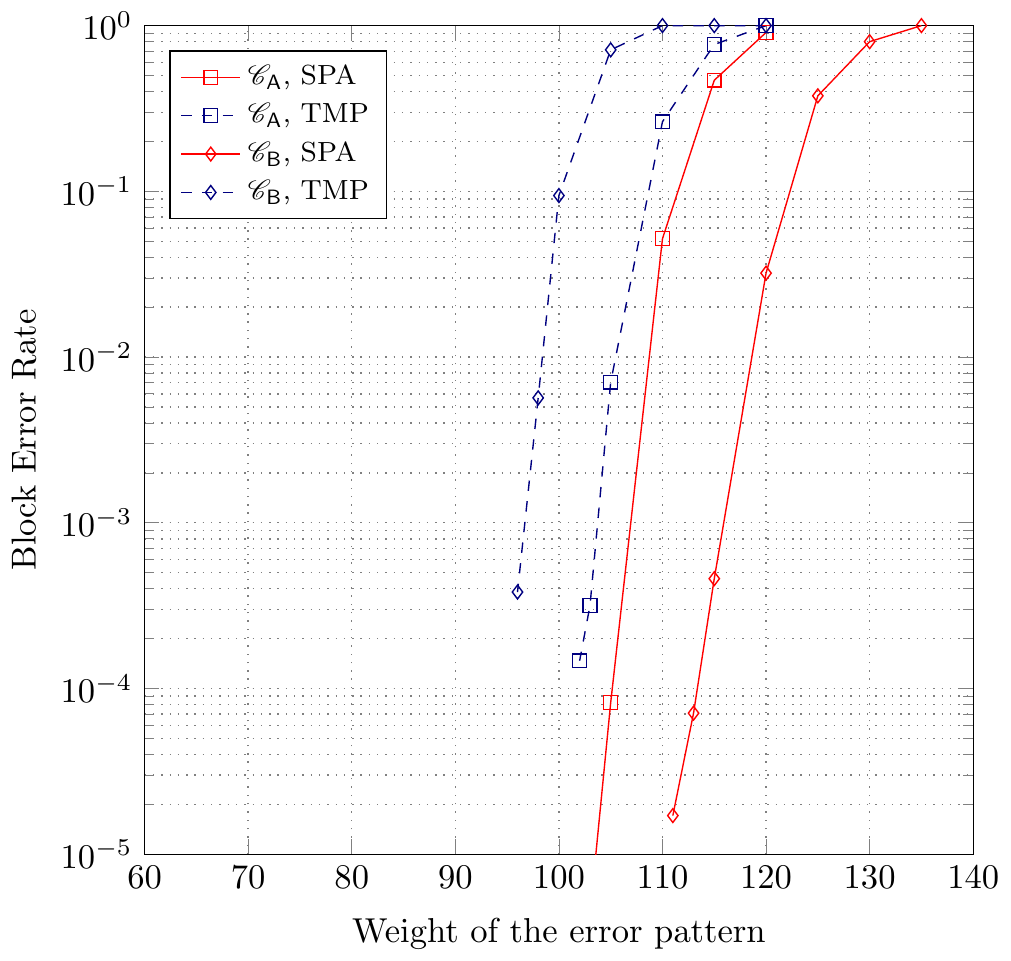}
	\caption{Block error rate for codes from the ensembles in Table~\ref{tab:thresholds} under \ac{SPA} and \ac{TMP} decoding with $100$ iterations.}
	\label{fig:9602SPA100it}
\end{figure}

Figure~\ref{fig:9602SPA100it} shows that the error correction performance of $\ens{B}$ significantly improves upon the performance of $\ens{A}$ under \ac{SPA} decoding.
For \ac{TMP} decoding we observe, that we can recover a big part of the loss with respect to $\ens{A}$.
The figure also shows that the decoding thresholds predicted by \ac{DE} (see Table~\ref{tab:thresholds}) give a good estimate of the error correction performance gains.
\section{Conclusions}\label{sec:conclusions}
\acresetall

In this paper, \ac{MP} decoding schemes for \ac{LDPC} codes with \acp{HWA}, like used e.g. in the post-quantum crypto NIST proposal LEDAcrypt, were considered. 
A new protograph-based decoding framework that allows to analyze and optimize \ac{MP} decoding schemes for \ac{LDPC} codes with \acp{HWA} was presented.
The new scheme uses a turbo-like principle to incorporate partial information about the errors that is available at the decoder and recovers most of the loss due to the error amplification due to the \ac{HWA}.
Decoding thresholds for the resulting code ensembles under \ac{SPA} and \ac{TMP} decoding were obtained using \ac{DE} analysis.
The results show, that the proposed decoding scheme improves upon basic decoding approach and has a similar performance as the \ac{MDPC} decoding approach with a significantly lower complexity.

% \bibliographystyle{IEEEtran}
% \bibliography{./latex-common/IEEEabrv,./latex-common/book}


% % \appendices

% \bibliography{./latex-common/IEEEabrv,./latex-common/book}

\begin{thebibliography}{10}
\providecommand{\url}[1]{#1}
\csname url@samestyle\endcsname
\providecommand{\newblock}{\relax}
\providecommand{\bibinfo}[2]{#2}
\providecommand{\BIBentrySTDinterwordspacing}{\spaceskip=0pt\relax}
\providecommand{\BIBentryALTinterwordstretchfactor}{4}
\providecommand{\BIBentryALTinterwordspacing}{\spaceskip=\fontdimen2\font plus
\BIBentryALTinterwordstretchfactor\fontdimen3\font minus
  \fontdimen4\font\relax}
\providecommand{\BIBforeignlanguage}[2]{{%
\expandafter\ifx\csname l@#1\endcsname\relax
\typeout{** WARNING: IEEEtran.bst: No hyphenation pattern has been}%
\typeout{** loaded for the language `#1'. Using the pattern for}%
\typeout{** the default language instead.}%
\else
\language=\csname l@#1\endcsname
\fi
#2}}
\providecommand{\BIBdecl}{\relax}
\BIBdecl

\bibitem{mceliece1978public}
R.~J. McEliece, ``{A Public-Key Cryptosystem Based on Algebraic Codes},''
  \emph{Deep Space Network Progr. Report}, vol.~44, pp. 114--116, 1978.

\bibitem{RSA1978}
R.~L. Rivest, A.~Shamir, and L.~Adleman, ``{A Method for Obtaining Digital
  Signatures and Public-Key Cryptosystems},'' \emph{Communications of the ACM},
  vol.~21, no.~2, pp. 120--126, Feb. 1978.

\bibitem{monico2000using}
C.~Monico, J.~Rosenthal, and A.~Shokrollahi, ``{Using Low Density Parity Check
  Codes in the McEliece Cryptosystem},'' in \emph{Proc. {IEEE} Int. Symp. Inf.
  Theory (ISIT)}, Sorrento, Italy, Jun. 2000, p. 215.

\bibitem{BaldiHWA}
M.~Baldi and F.~Chiaraluce, ``{Cryptanalysis of a New Instance of McEliece
  Cryptosystem based on QC-LDPC Codes},'' in \emph{IEEE Int. Symp. on Inf.
  Theory}, June 2007, pp. 2591--2595.

\bibitem{otmani2010cryptanalysis}
A.~Otmani, J.-P. Tillich, and L.~Dallot, ``{Cryptanalysis of two McEliece
  Cryptosystems based on Quasi-Cyclic Codes},'' \emph{Mathematics in Computer
  Science}, vol.~3, no.~2, pp. 129--140, 2010.

\bibitem{baldi2008new}
M.~Baldi, M.~Bodrato, and F.~Chiaraluce, ``{A New Analysis of the McEliece
  Cryptosystem Based on QC-LDPC Codes},'' in \emph{Int. Conf. on Security and
  Cryptogr. for Networks}.\hskip 1em plus 0.5em minus 0.4em\relax Springer,
  2008, pp. 246--262.

\bibitem{baldi2013optimization}
M.~Baldi, M.~Bianchi, and F.~Chiaraluce, ``{Optimization of the Parity-Check
  Matrix Density in QC-LDPC Code-Based McEliece Cryptosystems},'' in \emph{IEEE
  Int. Conf. on Comm. Workshops (ICC)}.\hskip 1em plus 0.5em minus 0.4em\relax
  IEEE, 2013, pp. 707--711.

\bibitem{BaldiLEDAcrypt}
M.~{Baldi}, A.~{Barenghi}, F.~{Chiaraluce}, G.~{Pelosi}, and P.~{Santini},
  ``{LEDAcrypt: Low-dEnsity parity-checkcoDe-bAsed cryptographic systems},''
  NIST PQC submission, March 2019, \url{https://www.ledacrypt.org/}.

\bibitem{Gallager63:LDPC}
R.~G. Gallager, \emph{Low-Density Parity-Check Codes}.\hskip 1em plus 0.5em
  minus 0.4em\relax Cambridge, MA, USA: M.I.T. Press, 1963.

\bibitem{Rudolph1967Majority}
L.~{Rudolph}, ``{A Class of Majority Logic Decodable Codes (Corresp.)},''
  \emph{IEEE Trans. Inf. Theory}, vol.~13, no.~2, pp. 305--307, April 1967.

\bibitem{Misoczki13:MDPC}
R.~Misoczki, J.~P. Tillich, N.~Sendrier, and P.~S. L.~M. Barreto,
  ``{MDPC-McEliece: New McEliece Variants from Moderate Density Parity-Check
  Codes},'' in \emph{IEEE Int. Symp. on Inf. Theory (ISIT)}, Istanbul, Turkey,
  Jul. 2013, pp. 2069--2073.

\bibitem{kobara2001semantically}
K.~Kobara and H.~Imai, ``{Semantically secure McEliece public-key
  cryptosystems-conversions for McEliece PKC},'' in \emph{Public Key
  Cryptography}, vol. 1992.\hskip 1em plus 0.5em minus 0.4em\relax Springer,
  2001, pp. 19--35.

\bibitem{Thorpe03:PROTO}
J.~Thorpe, ``{Low-Density Parity-Check {(LDPC)} Codes Constructed from
  Protographs},'' JPL IPN, Tech. Rep., Aug. 2003, 42-154.

\bibitem{Divsalar04:ARA}
A.~Abbasfar, K.~Yao, and D.~Disvalar, ``{Accumulate Repeat Accumulate Codes},''
  in \emph{Proc. IEEE Globecomm}, Dallas, Texas, Nov. 2004.

\bibitem{Divsalar09:ProtoJSAC}
D.~Divsalar, S.~Dolinar, C.~Jones, and K.~Andrews, ``{Capacity-approaching
  Protograph Codes},'' \emph{IEEE JSAC}, vol.~27, no.~6, pp. 876--888, August
  2009.

\bibitem{liva_pexit}
G.~Liva and M.~Chiani, ``{Protograph {LDPC} Code Design based on {EXIT}
  Analysis},'' in \emph{Proc. IEEE Globecomm}, Washington, US, Dec. 2007, pp.
  3250--3254.

\bibitem{bartz2019decoding}
H.~Bartz and G.~Liva, ``{On Decoding Schemes for the MDPC-McEliece
  Cryptosystem},'' in \emph{12th Int. ITG Conf. on Systems, Comm. and Coding
  (SCC)}.\hskip 1em plus 0.5em minus 0.4em\relax VDE, 2019, pp. 1--6.

\bibitem{lechner_analysis_2012}
G.~Lechner, T.~Pedersen, and G.~Kramer, ``Analysis and {{Design}} of {{Binary
  Message Passing Decoders}},'' \emph{{IEEE} Trans. Commun.}, vol.~60, no.~3,
  Mar. 2012.

\bibitem{yacoub2019protograph}
E.~B. Yacoub, F.~Steiner, B.~Matuz, and G.~Liva, ``{Protograph-Based LDPC Code
  Design for Ternary Message Passing Decoding},'' in \emph{SCC 2019; 12th
  International ITG Conference on Systems, Communications and Coding}.\hskip
  1em plus 0.5em minus 0.4em\relax VDE, 2019, pp. 1--6.

\bibitem{Chung01:DE}
S.-Y. Chung, G.~D. Forney, T.~J. Richardson, and R.~Urbanke, ``{On the Design
  of Low-Density Parity-Check Codes within $0.0045$ dB of the Shannon Limit},''
  \emph{{IEEE} Commun. Lett.}, vol.~5, no.~2, pp. 58--60, Feb 2001.

\bibitem{Jin2006:QDE}
H.~Jin and T.~Richardson, ``{A New Fast Density Evolution},'' in \emph{Proc.
  IEEE Inf. Theory Workshop (ITW)}, Punta del Este, Uruguay, March 2006, pp.
  183--187.

\bibitem{Liva2013:Proto_TWCOM}
P.~Pulini, G.~Liva, and M.~Chiani, ``{Unequal Diversity LDPC Codes for Relay
  Channels},'' \emph{{IEEE} Trans. Wireless Commun.}, vol.~12, no.~11, pp.
  5646--5655, Nov. 2013.

\bibitem{chung2001analysis}
S.-Y. Chung, T.~J. Richardson, and R.~L. Urbanke, ``{Analysis of Sum-Product
  Decoding of Low-Density Parity-Check Codes using a Gaussian Approximation},''
  \emph{{IEEE} Trans. Inf. Theory}, vol.~47, no.~2, pp. 657--670, Feb. 2001.

\end{thebibliography}

\end{document}